\documentclass[aps,superscriptaddress,twocolumn]{revtex4-2}
\usepackage{graphicx}
\usepackage{dcolumn}
\usepackage{bm}
\usepackage{xcolor}
\usepackage{amsmath}
\usepackage{amsfonts} 
\usepackage{appendix}
\usepackage{hyperref}
\usepackage{placeins}
\renewcommand{\vec}[1]{\mathbf{#1}}

\begin{document}
\title{Photon Phase-Space Dynamics in a Plasma Wakefield Accelerator}

\author{Neil Beri}
 \email{nebe@umich.edu}
 \affiliation{
 Gérard Mourou Center for Ultrafast Optical Sciences, University of Michigan, Ann Arbor, Michigan 48109, USA
}
 \author{John Palastro}
 \affiliation{Laboratory For Laser Energetics, University of Rochester,\\ Rochester, New York 14623, USA}
\author{Qian Qian}
\affiliation{
 Gérard Mourou Center for Ultrafast Optical Sciences, University of Michigan, Ann Arbor, Michigan 48109, USA
}
\author{Kyle Miller}
 \affiliation{Laboratory For Laser Energetics, University of Rochester,\\ Rochester, New York 14623, USA}

\author{Brandon K. Russell}
\affiliation{Department of Astrophysical Sciences, Princeton University, Princeton, New Jersey 08544, USA}

\author{Alexander G. R. Thomas}
 \email{agrt@umich.edu}
\affiliation{
 Gérard Mourou Center for Ultrafast Optical Sciences, University of Michigan, Ann Arbor, Michigan 48109, USA
}

\date{\today}

\begin{abstract}
Frequency up-shifting of laser light in a beam-driven plasma wakefield has the potential to provide high-intensity sources of short wavelength radiation. Simulations have demonstrated that a laser pulse can undergo large frequency shifts, limited only by the drive beam energy, when the plasma density is tailored to match the accelerating phase of the wake to the group velocity of the pulse. Here, we study the dynamical evolution of photons in the phase-space vicinity of the plasma wake-phase matching condition. Numerical calculations using a photon kinetic model are validated by direct comparison with full electromagnetic particle-in-cell simulations. These calculations form the basis of a linear theory of the photon dynamics which reveals several important results, including scalings for the properties of the witness pulse and a self-similar solution for the photon phase-space dynamics. One prediction of the theory is that the pulse can be compressed indefinitely with no lower bound on the duration. This predication suggests that photon acceleration can provide a novel source of sub-femtosecond, short wavelength radiation.  
\end{abstract}

\maketitle

\section{\label{sec:intro}Introduction}
\begin{figure*}[t!]
    \centering
    \includegraphics[width=\linewidth]{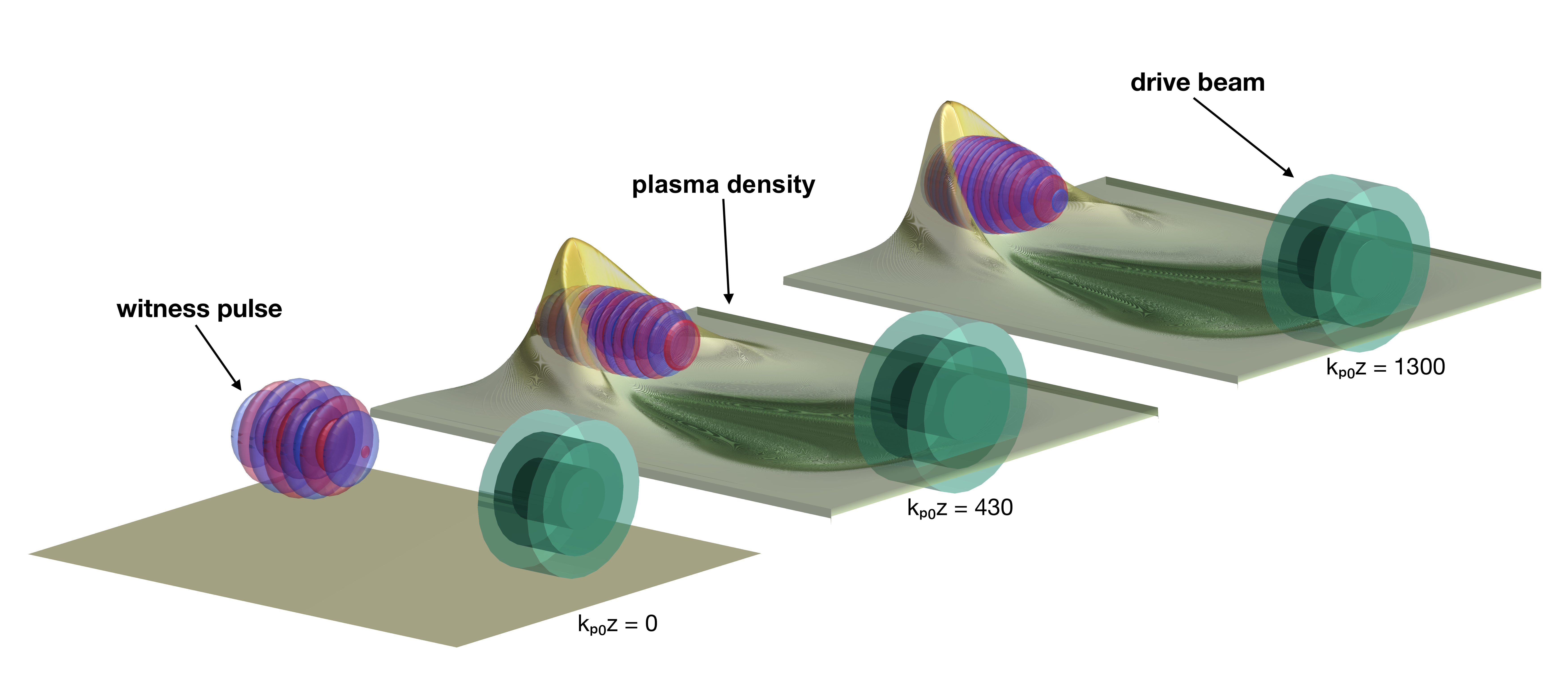}
    \caption{Depiction of PWPA. A charged particle beam excites a large amplitude plasma wave. A trailing laser pulse sits within the negative electron density gradient of the wave, where the associated traveling refractive index gradient causes it to upshift. The fields plotted are contours of the drive beam density (teal), contours of the electric field of the pulse (blue and red), and a surface representing the plasma density in the $xz$ plane (yellow-green). See column A of Table \ref{tab:simparams} for a complete list of simulation parameters.}
    \label{fig:pretty}
\end{figure*}

Photon acceleration refers to the modification of the frequency of light that travels in a spacetime-varying refractive index. Plasma wakefield photon acceleration (PWPA) \cite{wilks} is a mechanism in which the frequency of an ultrashort laser pulse is upshifted by the spacetime-varying refractive index of a charged particle beam or laser-driven wakefield. As the frequency of the pulse increases in a plasma, its group velocity approaches the vacuum speed of light, motivating the term "photon acceleration". Photon acceleration has been measured in experiments using plasma waves generated in laser wakefield acceleration \cite{Siders_PRL_1996,Murphy_POP_2006,Nie2018RelativisticStructure} and with a similar mechanism using an ionization front \cite{Dias_PRL_1997}.  Recently, it was demonstrated \cite{dePWPA} through simulations that tailoring of the plasma density in an ultra-relativistic electron beam-driven plasma wakefield prevents dephasing, allowing the acceleration of relativistically intense, optical pulses to extreme ultraviolet (XUV) wavelengths (see Fig.~\ref{fig:pretty}). Dephasing refers to the displacement of the witness pulse away from the accelerating phase of the wake due to the difference in velocities of the pulse and the drive beam. PWPA that uses the tailored density ramp is referred to as dephasingless. Simulations have demonstrated that the tailored plasma density profile \cite{Bulanov_PPR_1997,sprangle-hafizi-penano-etal-2001,guillaume-dopp-thaury-2015,bulanov-esirkepov-hayashi-etal-2016} can be achieved using a series of gas cells \cite{labdePWPA,modgasjet}.

Coherent, ultrashort pulsed XUV sources are desired for their ability to image the ultrafast dynamics of molecules and create ``molecular movies'' \cite{molmov}, diagnose the hydrodynamics of high-energy-density and warm-dense plasma \cite{hedpXUV}, image nanomaterials \cite{nanomaterials}, and perform lithography on microchips with nanoscale precision \cite{litho}. Although high laser intensities are available at optical wavelengths, current XUV sources, such as free electron lasers \cite{FEL} and high harmonic generation \cite{highharmonic}, have not yet reached relativistic intensities. XUV pulses generated from PWPA could potentially fill this intensity gap \cite{vortex}, providing a novel source of bright XUV light.

PWPA has several advantages such as its ability to compress, amplify, and preserve the structure of the witness pulse, motivating several studies of the mechanism. PWPA preserves the spatio-polarization structure of vector vortex pulses \cite{vortex} as they frequency upshift, amplify, and compress. Transition radiation from the drive beam entering the plasma could potentially self-seed the witness pulse, eliminating the need for an externally injected pulse \cite{selfseed}. 

In this work, we develop a time-dependent model for the Wigner distribution of the accelerating pulse using photon kinetic theory. We present both a procedure for numerically solving the photon kinetic equations in a plasma wakefield as well as a set of solutions to linearized equations. The photon kinetic model is used to show that dephasingless PWPA is stable. Scaling relationships for the XUV pulse parameters are derived and validated. These include the result that the pulse duration $\tau$ and bandwidth $\Delta\omega$ scale with the frequency shift $\omega/\omega_0$ as $\tau\sim\sqrt{\omega_0/\omega}$ and $\Delta\omega \sim\sqrt{\omega/\omega_0}$, and the pulse energy scales linearly as $U\sim{\omega/\omega_0}$. Here, $\omega$ is the pulse frequency after acceleration, and $\omega_0$ is the initial frequency of the pulse.

\section{\label{sec:pktwakefield}Photon Kinetic Theory in a Plasma Wakefield Photon Accelerator}
\subsection{\label{sec:nonlin} Full Photon Kinetic Theory}
Photon kinetic theory (PKT) \cite{Silva_PRE_1998,Silva_IEEE_2000,Mendonca} is an extension of geometric optics that describes the electromagnetic field as an ensemble of ``quasi-photons''  obeying a Hamiltonian given by their angular frequency $\omega$ with a momentum given by their wave vector $\vec{k}$. In this paper, quasi-photons will be referred to as photons for brevity. The local angular frequency as a function of the position and wave number can be obtained from the dispersion relation $D(\mathbf{k}, \omega, \mathbf{r}, t)=0$. An ensemble of quasi-photons is described by its phase-space density $N(\vec{r},\vec{k},t)$. Under certain conditions \cite{Silva_PRE_1998}, $N$ is approximately conserved along photon trajectories and can be interpreted as the photon density.
Here, we use a system of natural plasma units where $n_0=\omega_{p0}=c=m_e = e = \epsilon_0 = 1$, where $n_0$ is the reference electron density of the plasma and $\omega_{p0}=\sqrt{e^2n_0/(\epsilon_0 m_e)}$ is the reference plasma frequency. This means that time will be normalized to $\omega_{p0}^{-1}$, distance will be normalized to $k_{p0}^{-1} =c/\omega_{p0}$, and densities will be normalized to $n_0$. In these units, $\hbar = 1/(4\pi\alpha)$, where $\alpha$ is the fine structure constant.

For linear propagation in a relativistic, unmagnetized plasma, the dispersion relation is
\begin{equation}
    \label{plasmaD}
     D = 1 - \frac{k^2}{\omega^2} - \frac{\omega_p^2}{\omega^2}\;.
\end{equation}
The effective Hamiltonian of the quasi-photons in the plasma is therefore
\begin{equation}
    \omega(\vec{r},\vec{k},t) = \sqrt{|\vec{k}|^2 + \omega_p^2(\vec{r},t)} \approx  \left(|\vec{k}| + \frac{1}{2|\vec{k}|}\omega_p^2\right) \;,
     \label{plasmaH}
\end{equation}
where $\omega_p^2 = e^2 n/(\epsilon_0 m_e)$ is the squared electron plasma frequency with Lorentz invariant electron number density $n$, which is related to the electron number density in the laboratory frame by $n = n_{lab}/\gamma$, with $\gamma$ being the local, average Lorentz factor of the plasma electrons. The approximate form in Eq. \ref{plasmaH} is valid in the  underdense limit. Using Hamilton's equations of motion, one recovers the ray tracing equations of geometric optics,
\begin{equation}
    \label{canonicalEOM}
    \frac{d\mathbf{r}}{dt} = \frac{\partial\omega}{\partial\mathbf{k}}
    \;,\quad
    \frac{d\mathbf{k}}{dt} = -\frac{\partial\omega}{\partial\mathbf{r}}\;.
\end{equation}
When the following condition is satisfied, 
\begin{equation}
    \label{geoopts}
    \frac{2\pi}{\omega}\left| \frac{\partial}{\partial t}\ln(\eta) \right| + \frac{2\pi}{k}\left| \mathbf{\nabla} \ln(\eta) \right| \ll 1\;,
\end{equation}
where $\eta$ is the local index of refraction, the phase-space density $N$ is approximately conserved along photon trajectories \cite{Mendonca} and therefore evolves according to the transport equation
\begin{equation}
    \frac{\partial N}{\partial t}+\frac{d\vec{r}}{dt}\cdot\frac{\partial N}{\partial\vec{r}} +\frac{d\vec{k}}{dt}\cdot\frac{\partial N}{\partial \vec{k}} = 0\;.
\label{NVlasov} 
\end{equation}
The electric field $\vec{E}$ of the ensemble is related to the photon phase-space density via the Wigner distribution $\mathcal{F}$,
\begin{equation}
    \label{wig}
    \mathcal{F}(\mathbf{r}, \mathbf{k}, t) \equiv \int \mathbf{E}(\mathbf{r}-\frac{\mathbf{s}}{2}, t) \cdot \mathbf{E}^*(\mathbf{r}+\frac{\mathbf{s}}{2}, t) e^{i\mathbf{k}\cdot\mathbf{s}} d^3\mathbf{s}\;.
\end{equation}
An electromagnetic pulse can be interpreted as a coherent ensemble of quasi-photons with a small relative bandwidth centered at $k_\delta$. Its phase-space density is given by 
 \begin{equation}
    \label{Ndef}
    N(\mathbf{r}, \mathbf{k}, t) = \frac{1}{8\hbar}\frac{\partial D}{\partial\omega}\bigg|_{k=k_\delta} \mathcal{F}(\mathbf{r}, \mathbf{k}, t)\;.
\end{equation}
Since the wavenumber and frequency of an electromagnetic wave are approximately equal in an underdense plasma, the derivative in Eq. \eqref{Ndef} is given by $\partial D/\partial\omega |_{k=k_\delta} \approx 2/k_\delta$. The phase-space density can be integrated in either reciprocal or real space to yield the envelope or spectrum of the pulse as follows:
\begin{subequations}
    \label{intensities}
    \begin{align}
        |\mathbf{E}(\mathbf{r}, t)|^2 &= \left[ \frac{1}{8\hbar}\frac{\partial D}{\partial\omega}\bigg|_{k=k_\delta} \right]^{-1} \int N(\mathbf{r}, \mathbf{k}, t) \frac{d^3\mathbf{k}}{(2\pi)^3} \label{intensity1} \\
        |\hat{\mathbf{E}}(\mathbf{k}, t)|^2 &= \left[ \frac{1}{8\hbar}\frac{\partial D}{\partial\omega}\bigg|_{k=k_\delta} \right]^{-1} \int N(\mathbf{r}, \mathbf{k}, t) \frac{d^3\mathbf{r}}{(2\pi)^3}\;. \label{intensity2}
    \end{align}
\end{subequations}
The total energy of the electromagnetic field, including the magnetic energy, is 
\begin{equation}
    \label{Udef}
    U = 2\hbar\int\int\omega N(\mathbf{r}, \mathbf{k}, t)\frac{d^3\mathbf{k}}{(2\pi)^3}  d^3\mathbf{r}\;.
\end{equation}

When considering the evolution of an electromagnetic pulse in an electron beam-driven plasma wakefield, it is convenient to change coordinates from $z$ to the comoving coordinate $\xi \equiv z - \int_0^{t}v_d(\Tilde{t}) d\Tilde{t}$, with $\xi=0$ occurring at the head of the drive beam. For an ultrarelativistic drive beam, $v_d \rightarrow 1$, giving $\xi = z - t$. In this limit, it is also convenient to reparameterize the equations in terms of the $z$ coordinate of the drive beam since $v_d = dz/dt = 1$. The equations of motion in one spatial dimension become 
\begin{subequations}
    \label{EOMkxi}
    \begin{align}
        \frac{d\xi}{dz} &= -\frac{n}{2k^2} \label{EOMkxi1} \\
        \frac{dk}{dz} &= -\frac{1}{2k}\frac{\partial n}{\partial\xi}\;. \label{EOMkxi2}
    \end{align}
\end{subequations}

\begin{figure}[htbp]
    \centering
    \includegraphics[width=\linewidth]{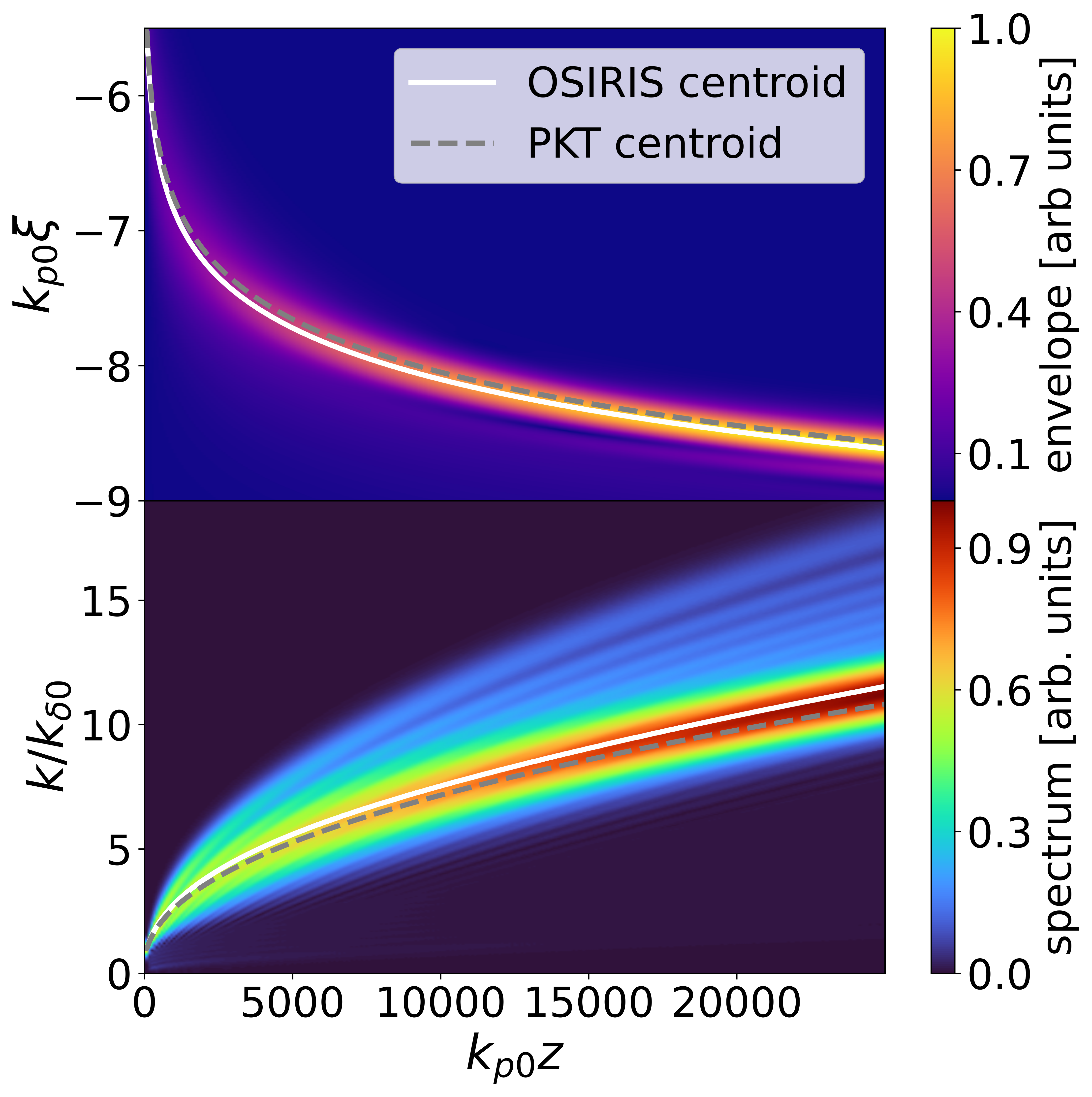}
    \caption{The plasma wake-phase matching conditions allow for continuous, monotonic frequency shift of the pulse. The plot tracks the centroid $(\xi_\delta, k_\delta)$ along with the spectrum and envelope of the pulse from an OSIRIS simulation. The predicted photon trajectory for the central photon (dashed gray) follows closely the trajectory of the pulse centroid calculated from the OSIRIS simulation (solid white). The initial central frequency was $k_{\delta0}=10$, initial pulse duration $\tau=1$, initial chirp $b_0=0$, drive beam density $n_d=0.8$, and beam length $L_d=1$. See column B of Table \ref{tab:simparams} for a complete list of simulation parameters.}
    \label{fig:disptraj}
\end{figure}

In Ref.~\cite{dePWPA}, it was shown that an initially tailored plasma density can eliminate dephasing, leaving the evolution of the drive beam as the main limitation to ``unlimited'' photon acceleration. The monotonically decreasing density profile $n_\delta(z) \equiv  n(\xi_\delta,z)$ ensures that the central photon of the pulse always remains at the point $(\xi_\delta, k_\delta)$ in phase space, where $\xi_\delta$ is the position in the wakefield where the density perturbation is zero with a negative gradient. The plasma wake-phase matching conditions that determine $\xi_\delta$, $k_\delta$, and $n_\delta$ are
\begin{subequations}
    \label{phasematching}
    \begin{align}
        \frac{d\xi_\delta}{dz} &= -\frac{n_\delta}{2k_\delta^2} \label{phasematchingxi}\\
        \frac{dk_\delta}{dz} &= A_d\frac{n_\delta^{3/2}}{2k_\delta} \label{phasematchingk} \\
        \frac{dn_\delta}{dz} &= -\frac{n_\delta}{2k_\delta^2}\left[ \frac{d\xi_\delta}{dn_\delta} \right]^{-1}\;, \label{phasematchingn}
    \end{align}
\end{subequations}
where the areal drive beam density is $A_d=n_dL_d$, $n_d$ is the number density of the drive beam, and $L_d$ is the length of the drive beam. The plasma wake-phase matching conditions were derived by modeling the wake behind the drive beam using the Akhiezer and Polovin 1D wake solutions \cite{AP} with an equilibrium density that varies with $z$. The wake within the beam was modeled using the method outlined in Ref.~\cite{wakeinbeam}. These equations are also written in the short or high-density limit where $n_d\gg1$ or $L_d\ll1$, leading to the maximum Lorentz factor in the wake $\gamma_m\approx1 + A_d^2/(2n_\delta)$. Figure \ref{fig:disptraj} shows agreement between the plasma wake-phase matching conditions for $(\xi_\delta, k_\delta)$ and high-resolution PIC simulations.  

Once the plasma density profile $n_\delta(z)$ is calculated as outlined in Ref.~\cite{dePWPA}, Eqs.~\eqref{EOMkxi} coupled with the Akhiezer and Polovin wake solutions can be numerically integrated to find the phase space trajectories of photons in a dephasingless PWPA. As an alternative to the analytic wakefield solutions, the plasma density and Lorentz factor can be calculated using a PIC or fluid simulation. The initial phase-space density is found by Wigner transforming the initial field of the pulse. This density is assigned as a weight to an ensemble of macroparticles distributed in the phase space. The evolution of the phase-space density is determined by interpolating the particle trajectories onto a rectangular grid. The time-dependent phase-space density can then be used to reconstruct the spectrum and envelope of the accelerated pulse and compare with direct finite-difference time domain electromagnetic calculations.

\subsection{\label{sec:linPKT}Linearized Photon Kinetic Theory}
Analytic solutions for the  photon phase-space trajectories can be found by linearizing the ray-tracing equations about the point $(\xi_\delta, k_\delta)$. Equation \eqref{NVlasov} was then solved using the method of characteristics. The kinetic equations describing the plasma wakefield photon accelerator offer a computational advantage over traditional field solvers, but they still require numerical integration for specific parameters. The linearized solutions presented here provide more general conceptual insight into the evolution of the accelerated pulse as well as scaling relationships between moments of the phase space density and various pulse and drive beam parameters.

\subsubsection{\label{sec:lintraj}Individual Photon Trajectories} 

To begin, Equations \eqref{EOMkxi} can be linearized about the point $(\xi_\delta, k_\delta)$ to yield
\begin{subequations}
    \label{linEOM}
    \begin{align}
        \frac{d\Delta\xi}{dz} &= -\frac{1}{2k_\delta^2}\frac{\partial n}{\partial\xi}\bigg|_{\xi_\delta}\Delta\xi + \frac{n_\delta}{k_\delta^3} \Delta k \label{linEOM1} \\
        \frac{d\Delta k}{dz} &= -\frac{1}{2k_\delta}\frac{\partial^2n}{\partial\xi^2}\bigg|_{\xi_\delta}\Delta\xi + \frac{1}{2k_\delta^2} \frac{\partial n}{\partial\xi}\bigg|_{\xi_\delta}\Delta k\;, \label{linEOM2}
    \end{align}
\end{subequations}
where $\Delta\xi \equiv \xi - \xi_\delta$ and $\Delta k \equiv k - k_\delta$. For the higher order terms to remain negligible, it is required that $A_d|\Delta\xi| \ll 1$ and $|\Delta k|/k_\delta \ll 1$. In the short or high-density limit, the derivatives may be calculated from the wakefield solutions as
    \begin{align}
        \frac{\partial n}{\partial\xi}\bigg|_{\xi_\delta} = -A_dn_\delta \;,\quad 
        \frac{\partial^2n}{\partial\xi^2}\bigg|_{\xi_\delta} = 2A_d^2n_\delta \;.
        \label{derivs}
    \end{align}
Note that all terms on the right-hand sides of Eqs.~\eqref{linEOM} are proportional to $n_\delta(z)$, a function that has no analytic form. However, since the right-hand side of Eqs.~\eqref{phasematchingk} is also proportional to $n_\delta$, the equations of motion can be reparameterized using the central wave number $k_\delta$ as the new independent variable:
\begin{subequations}
    \label{linEOMkdelta}
    \begin{align}
         \frac{d\Delta\xi}{dk_\delta} &= \frac{1}{k_\delta}\Delta\xi + \frac{2}{A_dk_\delta^2} \Delta k \label{linEOMkdelta1} \\
         \frac{d\Delta k}{dk_\delta} &= -2A_d\Delta\xi - \frac{1}{k_\delta}\Delta k\;. \label{linEOMkdelta2}
    \end{align}
\end{subequations}
Equation \eqref{linEOMkdelta1} contains information about the dispersive properties of the wakefield, with the first term on the right-hand side resulting from the spatial dependence of group velocity dispersion due to the variations in plasma density within the wake, and second term from the chromatic dependence of group velocity dispersion. Likewise, the first and second terms on the right-hand side of \eqref{linEOMkdelta2} arise from the spatial and chromatic dependencies of photon acceleration within the wake. 

Equations \eqref{linEOMkdelta} may be converted into a linear system with constant coefficients in $\Delta\xi$ and $\Delta k / k_\delta$  by introducing a function $\theta(k_\delta)=\ln(k_\delta/k_{\delta0})$, resulting in 
\begin{subequations}
    \label{linEOMkdelta_v2}
    \begin{align}
         \frac{d\Delta\xi}{d\theta} &= \Delta\xi + \frac{2}{A_d} \frac{\Delta k}{k_\delta} \label{linEOMkdelta1_v2} \\
         \frac{d}{d\theta}\left(\frac{\Delta k}{k_\delta}\right) &= -2A_d\Delta\xi - 2\frac{\Delta k}{k_\delta}\;.\label{linEOMkdelta2_v2}
    \end{align}
\end{subequations}
These equations describe a spiral to a stable equilibrium. Equations \eqref{linEOMkdelta_v2} were solved with the initial conditions $\Delta\xi(k_{\delta0}) = \Delta\xi_0$ and $\Delta k(k_{\delta0}) = \Delta k_0$, where $k_{\delta0}$ is the initial value of $k_\delta$. The solutions are given by
\begin{subequations}
    \label{lintraj}
    \begin{align}
        &\Delta\xi = \frac{R}{A_d}\sqrt{\frac{k_{\delta0}}{k_\delta}}\cos\left( \Theta \right) \label{lintraj1} \\
        &\Delta k = -k_{\delta0}R\sqrt{\frac{k_\delta}{k_{\delta0}}}\sin\left( \Theta + \psi\right)\;, \label{perturbedtrajk}
        \end{align}
\end{subequations}
where $\Theta(k_\delta) = \tfrac{\sqrt{7}}{2}  \theta(k_\delta) - \phi$, $\psi = \arctan\left(\tfrac{3}{\sqrt{7}}\right)$ and
\begin{subequations}
    \begin{align}
    &\tan(\phi) = \frac{\tfrac{3}{\sqrt{7}}A_d\Delta\xi_0+\tfrac{4}{\sqrt{7}}\tfrac{\Delta k_0}{k_{\delta0}}}{A_d\Delta\xi_0} \label{phi} \\
    &R^2 = \frac{16}{7}\left[ A_d^2\Delta\xi_0^2 + \frac{3}{2}A_d\Delta\xi_0\frac{\Delta k_0}{k_{\delta0}} + \frac{\Delta k_0^2}{k_{\delta0}^2} \right]\;. \label{R} 
\end{align}
\end{subequations}
The solutions show that, as the central frequency (or wavenumber) of the pulse upshifts, the photons within the vicinity of $(\xi_\delta, k_\delta)$ oscillate in position and wavenumber. 

\begin{figure}[htbp]
    \centering
    \includegraphics[width=\columnwidth]{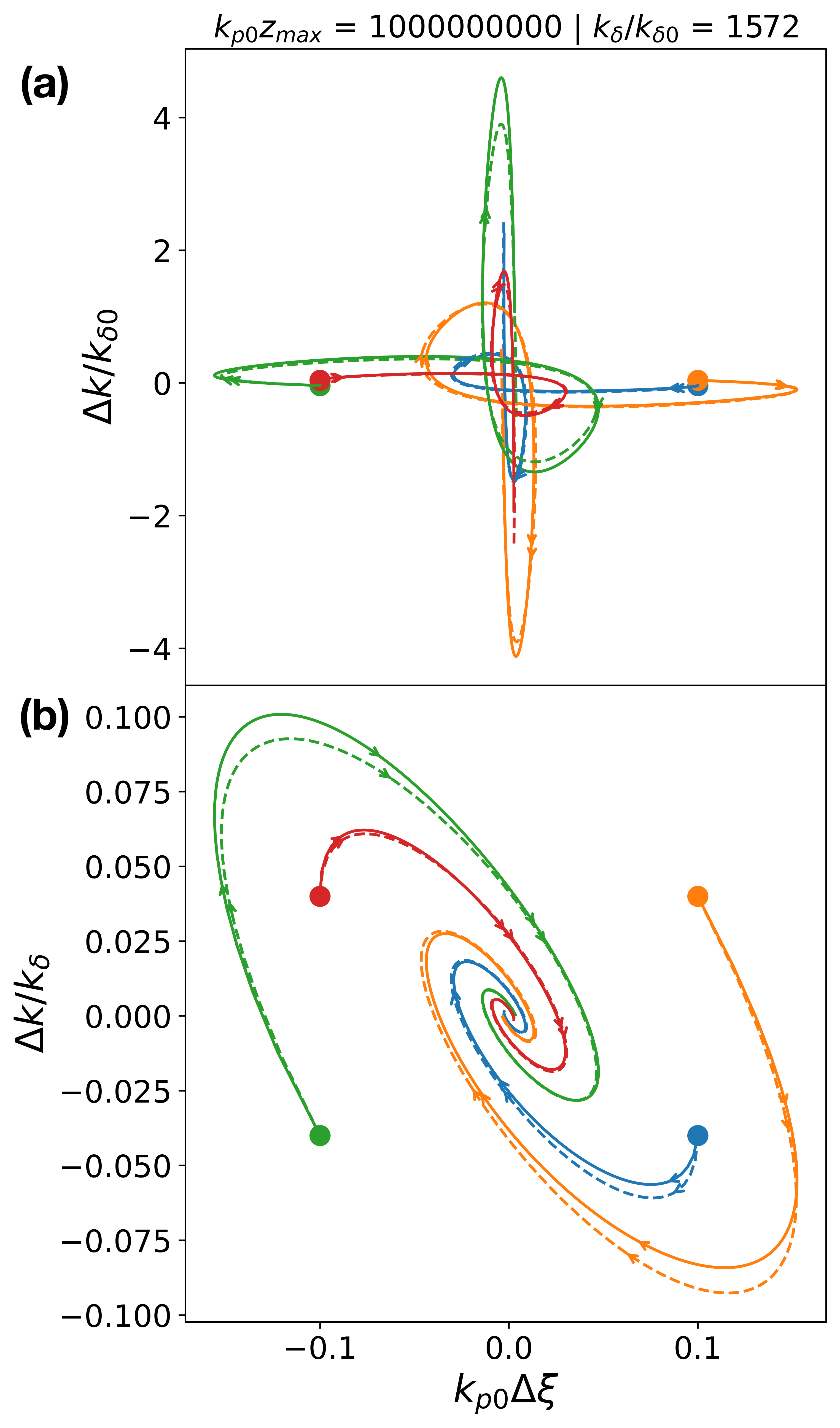}
    \caption{The photon phase space trajectories during PWPA calculated numerically (solid curves) are well approximated by the linearized solutions (dashed curves). Note that all trajectories spiral into the origin in $\Delta\xi$-$\Delta k/k_\delta$ space, as the origin is an attractor. The dots represent the initial conditions. (a) shows the trajectories in $\Delta\xi$-$\Delta k$ space, whereas (b) shows the trajectories in $\Delta\xi$-$\Delta k/k_\delta$ space. The parameters used were $n_d = 0.8$, $L_d = 1$, and $k_{\delta0} = 10$.}
    \label{fig:photontraj}
\end{figure}

Figure \ref{fig:photontraj} shows the linearized trajectories in phase space and compares them with the more exact, unlinearized numerical trajectories for a very large frequency shift. The linearized trajectories closely resemble the numerical curves, though the trajectories for which $|\Delta\xi|$ grows in the beginning develop more errors. Note that since $\Delta k$ grows slower than $k_\delta$, the relative deviation from the central frequency $\Delta k/k_\delta$ tends to zero.
Since the $\Delta\xi$ oscillations decrease during acceleration, and the $\Delta k$ oscillations grow slower than $k_\delta$, photons that satisfy $A_d|\Delta\xi| \ll 1$ and $|\Delta k|/k_\delta \ll 1$ before acceleration will satisfy said conditions for the entirety of the acceleration process. The phase space orbits are therefore also stable in space, as $\lim_{z\to\infty}\Delta\xi = 0$. As a result, in the phase-space $(\Delta\xi,\Delta k/k_\delta)$ the fixed point ($\xi_\delta,k_\delta$) is an attractor. 

\subsubsection{\label{sec:photondist}Photon Phase Space Density}
The method of characteristics can be used to analytically model the phase-space density of the accelerated pulse. For an initially chirped pulse with a Gaussian temporal profile, the electric field envelope can be expressed as $E_x(\Delta\xi, z=0) = E_0 \exp\left[ -2\Delta\xi^2/\tau^2 + i\left( k_{\delta0}\Delta\xi + \frac{1}{2}b_0\Delta\xi^2 \right) \right]$. The initial phase-space density is then
\begin{align}
    N_0(\Delta\xi_0, \Delta k_0) =& \frac{\sqrt{\pi} |E_0|^2\tau}{4 \hbar k_{\delta0}} \exp\left[ -\frac{\Delta\xi_0^2}{(\tau/2)^2} \right]  \label{N0}\\
    \times&\exp\left[ - \frac{\left(\Delta k_0 - b_0\Delta\xi_0\right)^2}{(2/\tau)^2}  \right]\;. \notag
\end{align}
Note that for this definition of the duration $\tau$, the full-width-at-half-maximum of the light intensity would be $\tau_{\text{FWHM}} = \sqrt{\ln(2)}\tau$. Equations \eqref{lintraj} were recast with their initial conditions serving as the dependent variables, and the subsequent expressions were substituted into \eqref{N0} to yield the time-dependent phase-space density. The resulting phase-space density expressed in terms of its moments is
\begin{align}
    N(\Delta\xi, \Delta k, k_\delta) =& \frac{\sqrt{\pi} |E_0|^2\tau}{4 \hbar k_{\delta0}} \exp\left[ -\frac{\Delta\xi^2}{2\langle\Delta\xi^2\rangle}\right] \label{Ndepwpa} \\ \times&\exp\left[-\frac{\left(\Delta k - b\Delta\xi\right)^2}{2\langle\Delta k^2\rangle}\right] ,\notag
\end{align}
where $b=\langle\Delta\xi\Delta k\rangle/\langle\Delta\xi^2\rangle$ is the linear chirp coefficient, 
and the full expressions for the moments are given in  Appendix \ref{appendixmoments}. This definition of chirp is equivalent to the gradient of the instantaneous frequency of the pulse. The resulting envelope and spectrum, obtained using Eqs.~\eqref{intensities}, are
\begin{subequations}
    \label{specenv}
    \begin{align}
        \left|\frac{E(\Delta\xi, k_\delta)}{E_0}\right|^2 &= \sqrt{\frac{\tau^2/8}{\langle\Delta\xi^2\rangle}}\frac{k_\delta}{k_{\delta0}}\exp\left[ -\frac{\Delta\xi^2}{2\langle\Delta\xi^2\rangle} \right] \label{envelope} \\
        \left|\frac{\hat{E}(\Delta k, k_\delta)}{E_0}\right|^2 &= \sqrt{\frac{\tau^2/8}{\langle\Delta k^2\rangle}}\frac{k_\delta}{k_{\delta0}}\exp\left[ -\frac{\Delta k^2}{2\langle\Delta k^2\rangle} \right] \;.\label{spectrum} 
    \end{align}
\end{subequations}
For the equations above to be accurate, the following conditions must be satisfied: $A_d\tau \ll 1$ and $k_\delta\tau \gg 1$, or $A_d \ll 1/\tau \ll k_\delta$. However, PIC simulations presented in the following section show that the linearized theory remains an adequate estimate even when these conditions are not met.

The moments of the phase-space density exhibit qualitative behavior that is similar to the photon trajectories they were derived from. The pulse duration (proportional to $\sqrt{\langle\Delta\xi^2\rangle}$) oscillates with the frequency shift with an ever-growing period and an amplitude that decays as $\sqrt{k_{\delta0}/k_\delta}$. The bandwidth (proportional to $\sqrt{\langle\Delta k^2\rangle}$), oscillates with the same period as the duration but with an amplitude that grows as $\sqrt{k_{\delta}/k_{\delta0}}$, a requirement imposed by the Fourier transform limit. These moments lead to the scalings $\sqrt{\langle\Delta\xi^2\rangle}\sim\sqrt{k_{\delta0}/k_\delta}$ and $\sqrt{\langle\Delta k^2\rangle}\sim\sqrt{k_{\delta}/k_{\delta0}}$. The theory predicts that the pulse duration approaches zero, so arbitrarily short pulses can be produced through PWPA with the evolution of the drive beam being the only limiting factor. The covariance of the phase-space density oscillates with a fixed amplitude, and the accelerating pulse is transform-limited when $\langle\Delta\xi\Delta k\rangle = 0$. As is true for the phase-space density of any linearly chirped Gaussian pulse, the generalized variance $|\Sigma|$ is constant such that $\sqrt{|\Sigma|}=\sqrt{\langle\Delta\xi^2\rangle\langle\Delta k^2\rangle - \langle\Delta\xi\Delta k\rangle^2} = \frac{1}{2}$. The evolution of the phase-space density can therefore be described by three transformations: contraction along the $\xi$-axis, expansion along the $k$-axis, and clockwise rotation about $(\xi_\delta, k_\delta)$. The $\xi$ contraction and $k$ expansion lead to spatial compression and spectral broadening of the pulse. The rotation of the phase-space density causes the pulse to either spatially compress or stretch, depending on the phase of the rotation. However, the rotation period grows continuously in time, so the contraction in $\xi$ and expansion in $k$ become the dominant transformations for large frequency shifts. 

Ignoring the slowly varying sinusoidal factors, the chirp of the pulse $b$ increases linearly with the frequency shift. Equation \eqref{envelope} predicts that the amplitude of the electric field grows as $(k_\delta/k_{\delta0})^{3/4}$, such that the total energy of the pulse is $U = \sqrt{\pi}|E_0|^2\tau/4 \times k_\delta/k_{\delta0}$. The total energy grows linearly with the frequency shift, meaning that the accelerated pulse is not only frequency upshifted and compressed, but also amplified. The linear dependence of the energy on the frequency shift reflects the fact that photon acceleration is a process that increases the energy $\omega \approx k$ of each individual photon by upshifting their frequencies while conserving photon number. 

\section{\label{sec:comp}Theory Validation}
The phase-space density predicted by PKT was tested by comparing to phase-space densities calculated from the Wigner distribution of the transverse electric field of the accelerating pulse from PIC simulations. The simulations were conducted in 1D using the dispersion free \cite{fei} electromagnetic solver in OSIRIS 4.0 \cite{osiris}. Details of all simulations are given in Appendix \ref{sec:simparams}.

Figure \ref{fig:wigcontour} demonstrates the ability of PKT to reproduce the phase space density from the same simulation. The PKT simulation was performed by numerically solving Eqs.~\eqref{EOMkxi} (using an RK45 algorithm) for a grid of 256 by 256 trajectories with initial conditions centered at $(\xi_{\delta0}, k_{\delta0})$ and interpolating their phase space densities onto a grid using a histogram. 
\begin{figure}[h!]
    \centering
    \includegraphics[width=\linewidth]{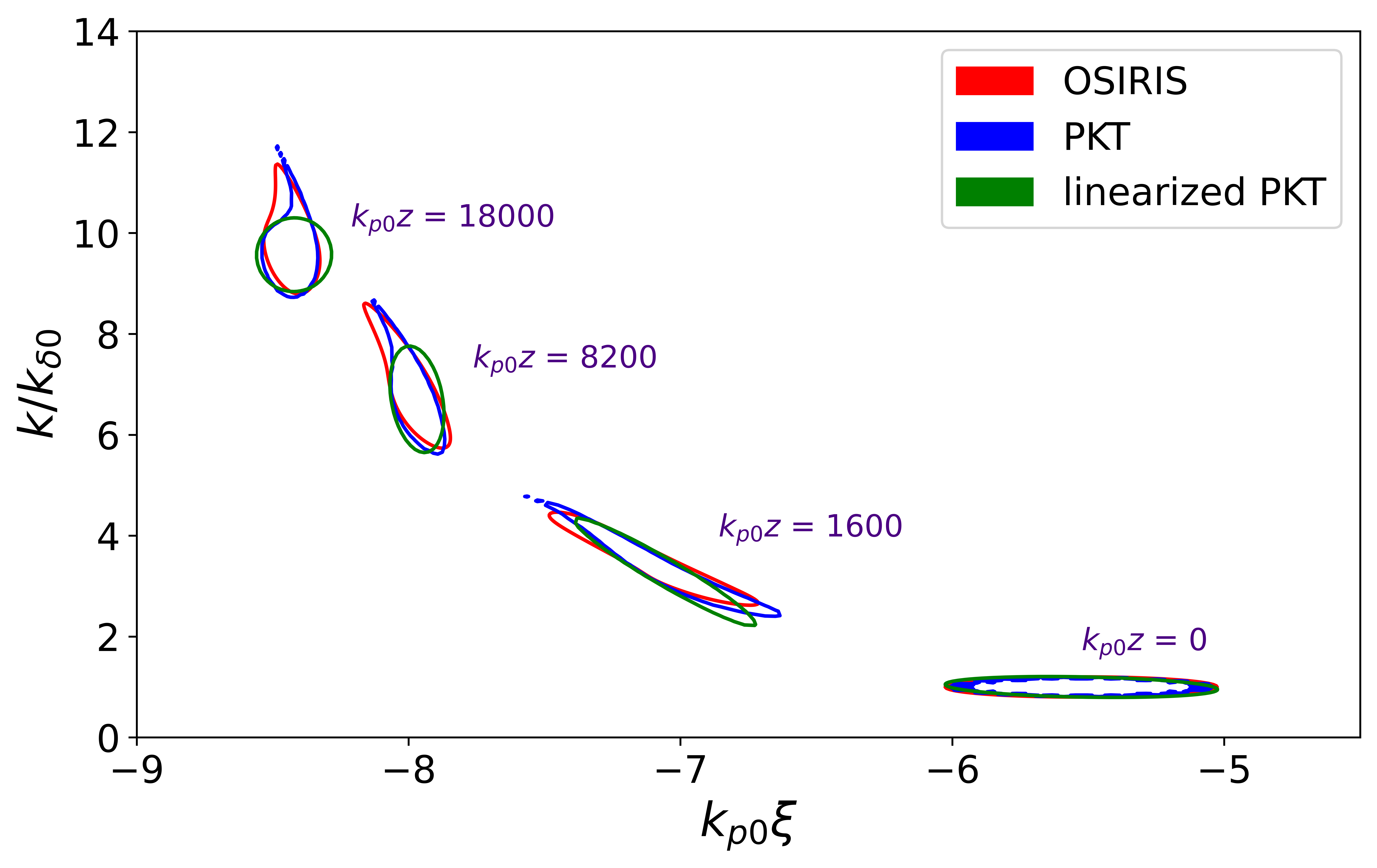}
    \caption{The phase-space density of the accelerating pulse contracts in space, expands in wavenumber, and rotates in phase-space. Contours of the photon phase-space density at $\exp(-1)$ times the maximum are plotted. The photon kinetic model is able to reproduce the phase-space density of the OSIRIS simulation. The linearized model is consistent with the OSIRIS simulation, though it breaks down for large frequency shifts when the a high-frequency tail forms in the OSIRIS simulation. The parameters used were $k_{\delta0}=10$, $\tau=1$, $n_d=0.8$, $b_0=0$, and $L_d=1$. See column B of Table \ref{tab:simparams} for a complete list of simulation parameters.}
    \label{fig:wigcontour}
\end{figure}
As the pulse is accelerated, its phase-space density rotates about the centroid, compressing in real space and expanding in reciprocal space. Photons that reach low $\Delta\xi$ and low $\Delta k$ have a much slower group velocity than the centroid, as they have low frequencies and are in a region of high plasma density. These photons thus spend more time in a region with a stronger accelerating gradient and are accelerated faster than the linear theory predicts. The rapidly accelerating photons form a high-frequency `tail' on the phase-space density that is captured by the PKT simulation. The linearized theory does not account for this phenomenon, however, as it treats the forces as symmetric about $\xi_\delta$. It can be seen in Fig.~\ref{fig:wigcontour} that as the `tail' develops, the phase-space density from linearized PKT becomes a worse approximation for the full solution. This is expected as $A_d=0.8$ and $\tau=1$, so the condition $A_d\tau\ll1$ is not satisfied. 

\begin{figure}[htbp]
    \centering
    \includegraphics[width=\linewidth]{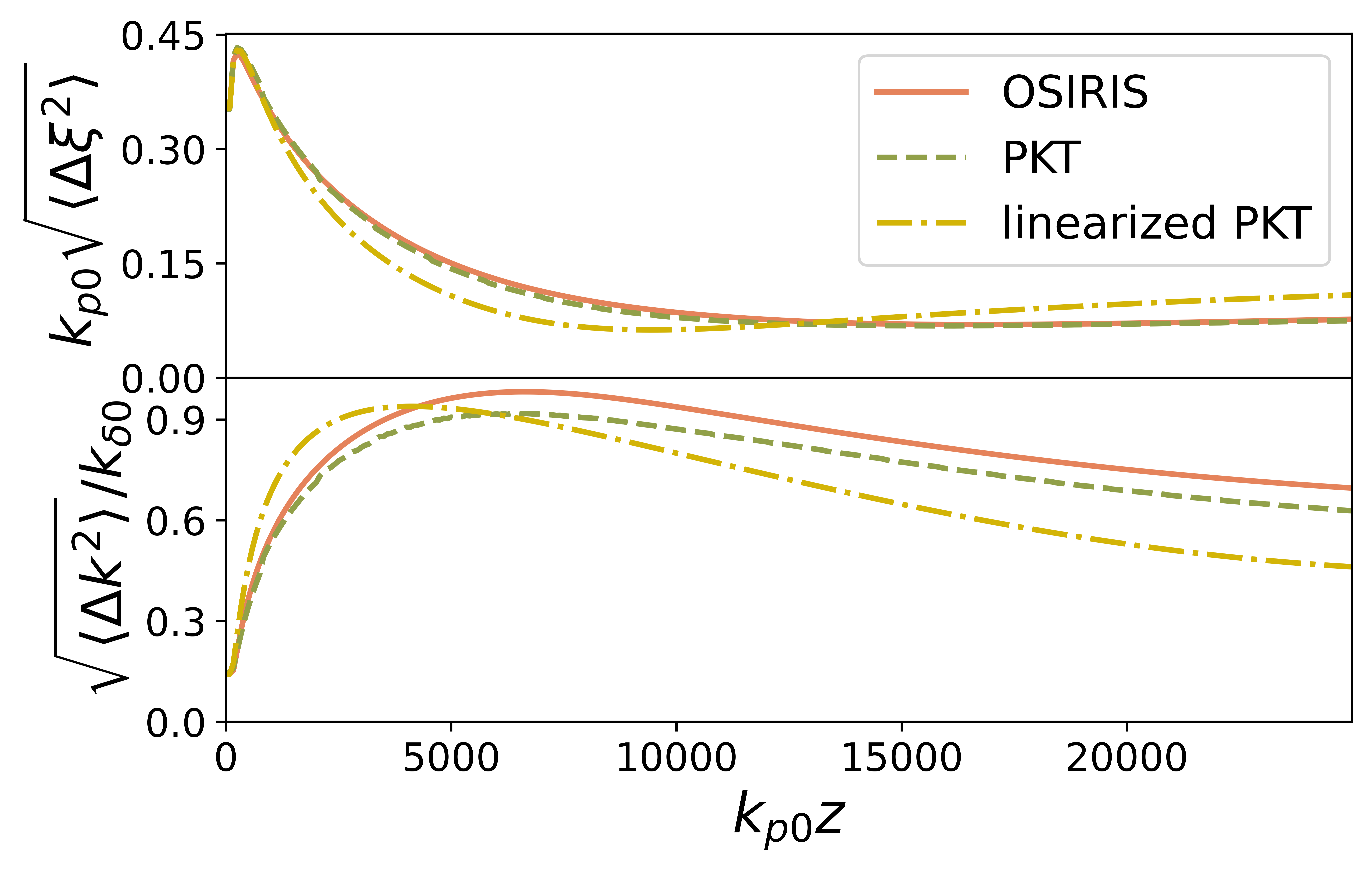}
    \caption{As the pulse is accelerated, it compresses and spectrally broadens. The plot depicts moments of the phase space densities from Fig.~\ref{fig:wigcontour} calculated by fits to a Gaussian. The linearized theory is able to predict the duration and bandwidth of the pulse as calculated from and OSIRIS simulation. Due to the phase-space rotation, the pulse initially stretches before compressing. The parameters used were $k_{\delta0}=10$, $\tau=1$, $n_d=0.8$, $b_0=0$, and $L_d=1$. See column B of Table \ref{tab:simparams} for a complete list of simulation parameters.}
    \label{fig:moments}
\end{figure}

Despite the condition for linearity not being satisfied, the linearized solutions are still an adequate approximation for the bulk of the distribution. To demonstrate this, the pulse duration and bandwidth of pulse are presented in Fig.~\ref{fig:moments}. In an effort to exclude the effect of the high-frequency `tails' on the OSIRIS and PKT distributions, the moments were calculated by fitting the spectra and envelopes of the electric fields to Gaussian functions. It can be seen from this plot that the  PKT accurately reproduces the result from the full electromagnetic solver and that the linear theory is in reasonable agreement, even though the conditions are far from ideal.

\section{\label{sec:selfsim} Self-Similar Solutions to the Linear Theory}
The evolution of the phase-space density of a Gaussian pulse is complex, undergoing phase space rotations as well as spatial contractions and spectral expansions. In order to isolate the rotational motion from the contraction in $\xi$ and expansion in $k$, it is useful to analyze the photon motion in a set of scaled coordinates $\chi_1$ and $\chi_2$:
\begin{align}
    \label{chichip}
    \chi_1 = \sqrt{\frac{k_\delta}{k_{\delta0}}}A_d\Delta\xi \;,\quad \chi_2 = \sqrt{\frac{k_{\delta0}}{k_\delta}}\frac{\Delta k}{k_{\delta0}}\;.
\end{align}
The rescalings in \eqref{chichip} lead to the photon trajectories oscillating with constant amplitude. Furthermore, this choice leads to conditions for small perturbations $|\chi_{10}| \ll 1$ and $|\chi_{20}| \ll 1$ that don't depend on system parameters. The equations of motion and trajectories in these coordinates become
\begin{subequations}
    \label{EOMchi}
    \begin{align}
        \frac{d\chi_1}{d\theta} &= \frac{3}{2}\chi_1 + 2\chi_2 \label{EOMchi1} \\
        \frac{d\chi_2}{d\theta} &= -2\chi_1 - \frac{3}{2}\chi_2 \;.\label{EOMchi2}
    \end{align}
\end{subequations}
\begin{subequations}
    \label{lintrajchi}
    \begin{align}
        \chi_1(k_\delta) &= R\cos(\Theta) \label{lintrajchi1} \\
        \chi_2(k_\delta) &= -R\sin(\Theta + \psi)\;. \label{lintrajchi2} 
    \end{align}
\end{subequations}
\begin{figure}[htbp]
    \centering
    \includegraphics[width=\linewidth]{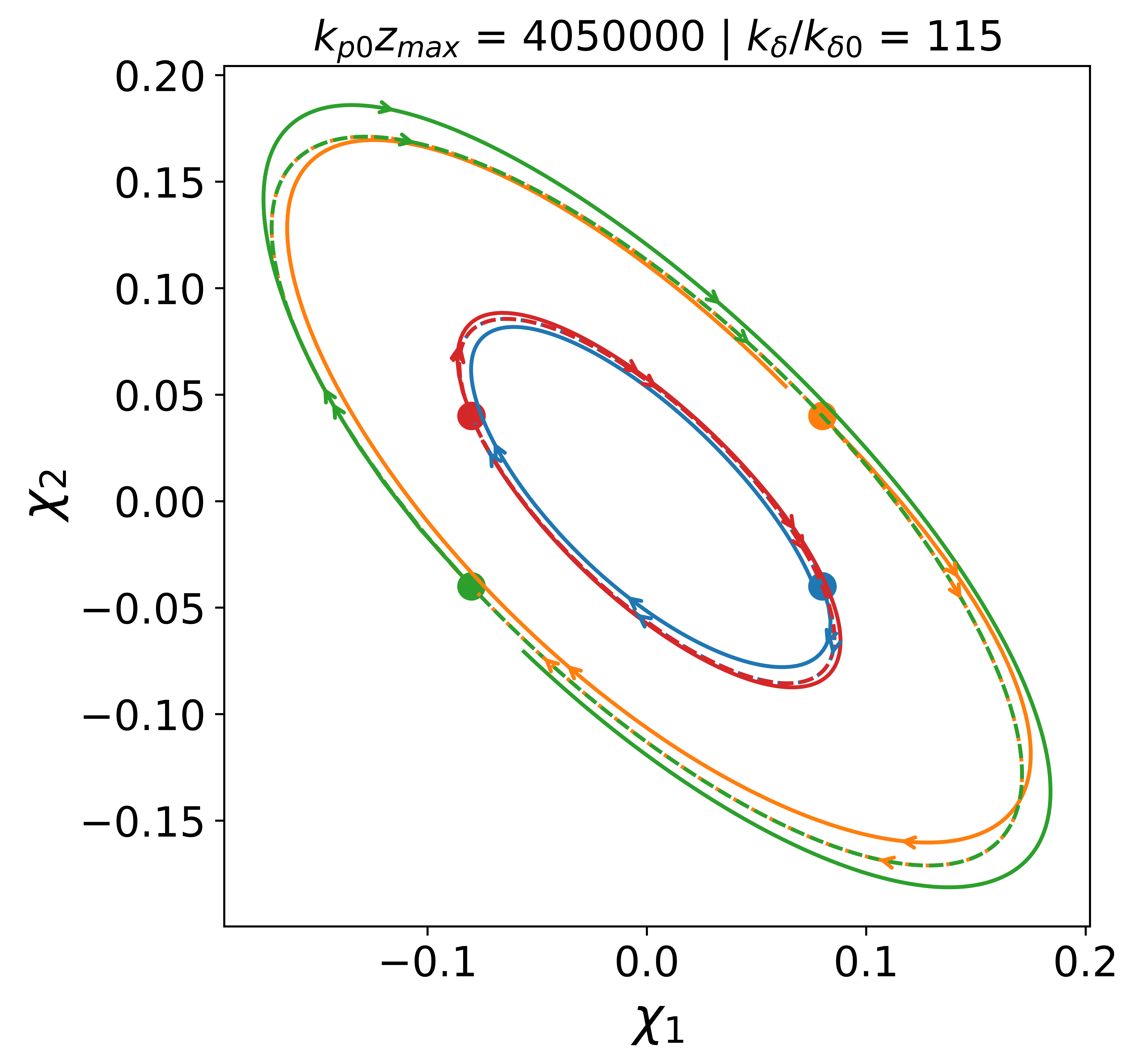}
    \caption{Photon phase space trajectories in scaled $\chi_1$-$\chi_2$ phase-space during PWPA calculated numerically (solid curves) and through the linearized solutions (dashed curves). In the scaled phase-space, the photon trajectories form closed elliptical orbits. The dots represent the initial conditions. The trajectories shown use the same parameters as Fig.~\ref{fig:photontraj}. The parameters used were $n_d = 0.8$, $L_d = 1$, and $k_{\delta0} = 10$.}
    \label{fig:photontrajchi}
\end{figure}
\begin{figure*}[t!]
    \centering
    \includegraphics[width=\linewidth]{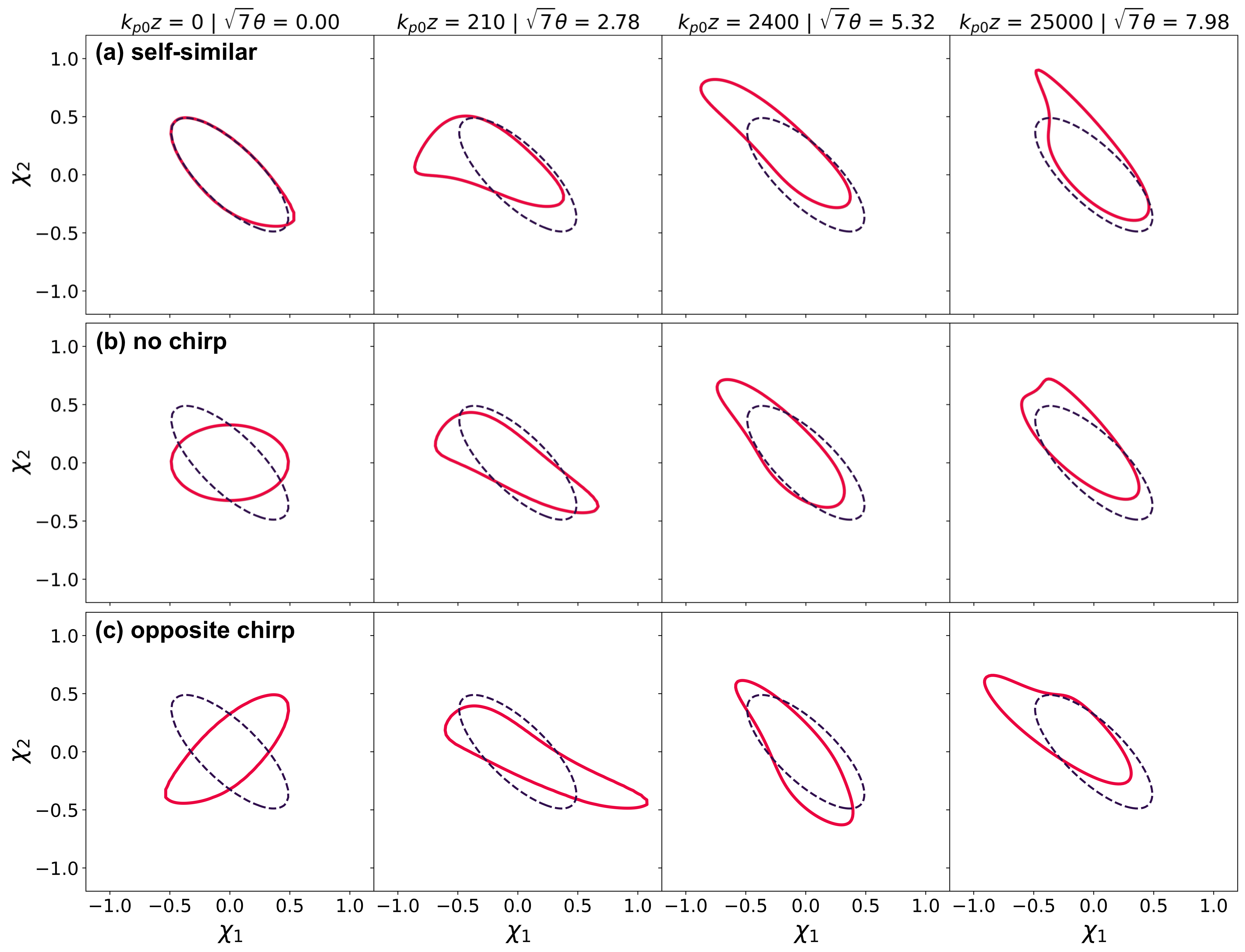}
    \caption{Evolutions of the exp(-1) contours of (a) a pulse (red) initialized as a self-similar distribution (dashed black) and (b) a pulse initialized with no chirp but with the same duration as the self-similar solution, and (c), a pulse initialized with the opposite chirp as the self-similar solution but with the same duration. The self-similar pulse remains nearly stationary during acceleration, while the other pulses undergo phase-space rotations that slow down near the self-similar distribution.The pulse parameters used were $\tau=2.06$, $n_d=2.37$, $L_d=0.2$, and $b_0=-1.06,\; 0,\;1.06$. See column C of Table \ref{tab:simparams} for a complete list of simulation parameters.}
    \label{fig:wignerchi}
\end{figure*}
\begin{figure*}[t!]
    \centering
    \includegraphics[width=\linewidth]{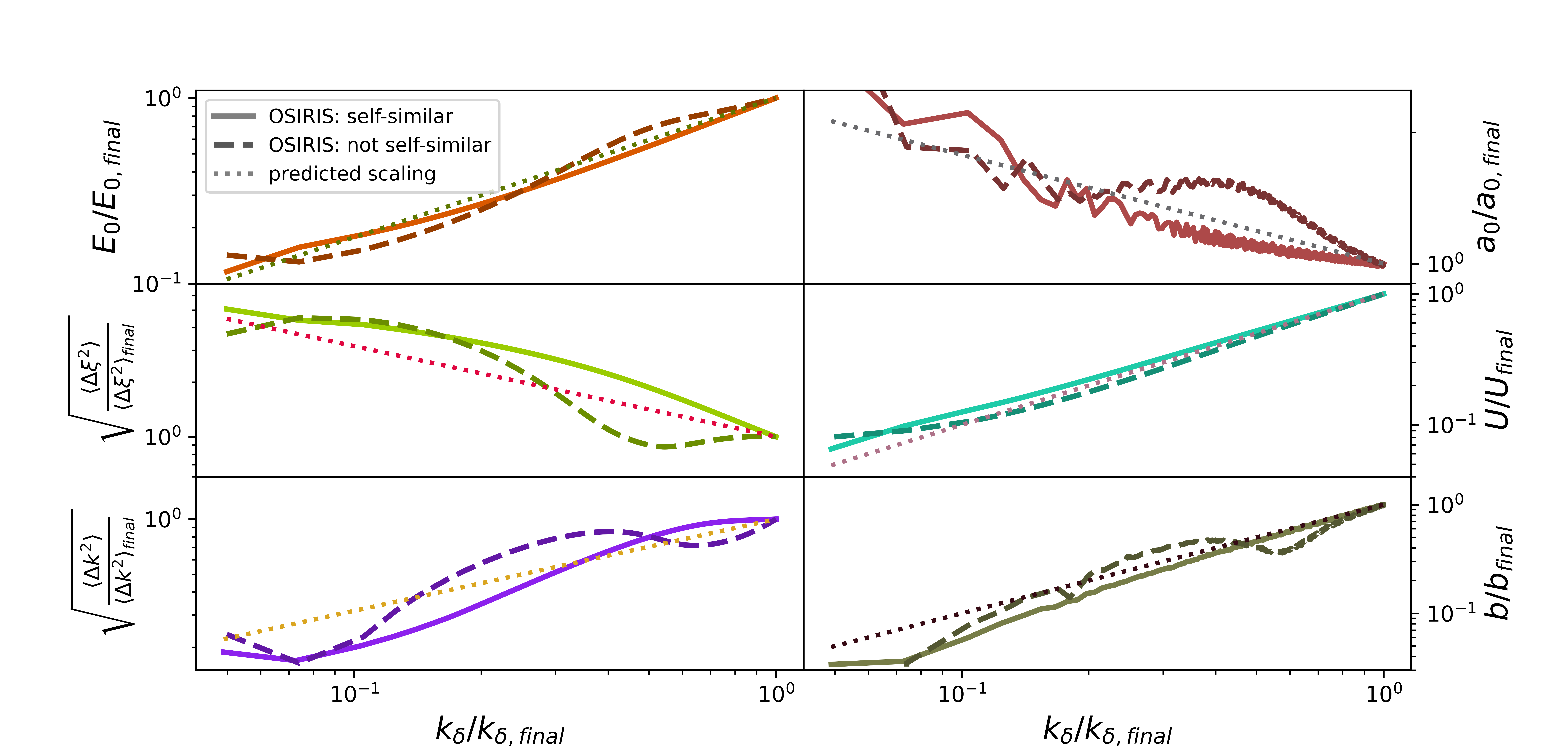}
    \caption{Comparison between OSIRIS simulations and theoretical predictions of scaling of pulse parameters with frequency shift. The parameters of the self-similar pulse closely follow the predicted scaling, whereas the parameters of the non-self-similar pulse oscillate about the predicted scaling. The simulation parameters used were $\tau=2.06$, $n_d=2.37$, $L_d=0.2$, and $b_0=-1.06$ (self-similar) and $b_0=1.06$ (not self-similar). See column C of Table \ref{tab:simparams} for a complete list of simulation parameters.}
    \label{fig:scaling}
\end{figure*}

It is now evident that the trajectories in $\chi_1$-$\chi_2$ space trace an ellipse with its major axis along the line $\chi_2=-\chi_1$. The equation for the ellipse is 
\begin{align}
    \label{ellipse}
    \frac{4}{\sqrt{7}}\chi_1^2 + \frac{6}{\sqrt{7}}\chi_1\chi_2 + \frac{4}{\sqrt{7}}\chi_2^2 = \frac{\sqrt{7}}{4}R^2 \equiv \epsilon_\gamma\;.
\end{align}
 Note that the coefficients are chosen to resemble the Courant-Snyder (Twiss) parameters that describe the phase-space trajectories of charged particles in accelerators. The area of the ellipse in $\chi_1$-$\chi_2$ space (aside from a factor of $\pi$) is $\epsilon_\gamma=\sqrt{7}R^2/4$, which is analogous to the emittance in accelerator physics. The major axis of the ellipse is skewed by an angle of $-\pi/4$, and it has an eccentricity of $\sqrt{2\sin|\psi|/(1+\sin|\psi|)}=\sqrt{6/7}$. Furthermore, the phase-space density in the scaled coordinates no longer compresses or spectrally broadens with no bound. Instead, each element of the covariance matrix of the phase-space density is $2\pi$-periodic in $\sqrt{7}\theta(k_\delta)$. Sample photon trajectories in $\chi_1$-$\chi_2$ space are shown in Fig.~\ref{fig:photontrajchi}.

It can thus be reasoned that, if the Gaussian pulse is initialized such that the contours of its phase-space density are ellipses of the form \eqref{ellipse}, the phase-space density will be stationary in $\chi_1$-$\chi_2$ space. In $\Delta\xi$-$\Delta k$ space, the phase-space density would thus strictly compress spatially and broaden spectrally with no oscillating factors in its covariance matrix. By requiring the contours of the  phase-space density to coincide with photon trajectories and enforcing that $|\Sigma|=1/4$ (in unscaled coordinates), we find the moments of the stationary distribution to be
\begin{subequations}
    \label{matchmoments}
    \begin{align}
        \langle\chi_1^2\rangle = \langle\chi_2^2\rangle &= \frac{2A_d}{\sqrt{7}k_{\delta0}} \label{varchi}\\
        \langle\chi_1\chi_2\rangle &= -\frac{3A_d}{2\sqrt{7}k_{\delta0}}\;. \label{covarchi}
    \end{align}
\end{subequations}
To achieve this self-similar solution, the pulse must have the correct initial conditions for a particular wakefield. By comparing Eqs.~\eqref{matchmoments} to the initial moments of the general Gaussian phase-space density Eq.~\eqref{N0}, the required initial chirp and pulse duration for the self-similar solution are found to be
\begin{subequations}
    \label{pulsematching}
    \begin{align}
        \tau &= \frac{4}{7^{1/4}\sqrt{A_dk_{\delta0}}} \label{taumatching} \\
        b_0 &= -\frac{3}{4}A_dk_{\delta0}. \label{bmatching}
    \end{align}
\end{subequations}
For a pulse initialized with this duration and chirp, the phase-space density is
\begin{align}
    \label{Nmatch}
    N(\chi_1, \chi_2, k_\delta) =& \frac{\sqrt{\pi} |E_0|^2\tau}{4 \hbar A_d} \\
    \times&\exp\left[ -\frac{4k_{\delta0}}{\sqrt{7}A_d}\left( \chi_1^2 + \frac{3}{2}\chi_1\chi_2 + \chi_2^2 \right)\right]\;, \notag
\end{align}
which is notably independent of $k_\delta$. The coordinate transformation presented in Eq.~\eqref{chichip} is thus a similarity transformation which reduces the number of independent variables in the system from three ($\Delta\xi$, $\Delta k$, $k_\delta$) to two ($\chi_1$, $\chi_2$). As a result, the phase-space density in Eq.~\eqref{Nmatch} is a self-similar solution to the linearized photon kinetic equation with similarity parameters $\chi_1$ and $\chi_2$.  The phase-space density \eqref{Nmatch} remains stationary in $\chi_1$-$\chi_2$ space, and phase-space densities with different $\tau$ and $b_0$ will oscillate about the self-similar density with increasing $\theta(k_\delta)$.

As observed in the OSIRIS simulations presented in Fig.~\ref{fig:wignerchi}, the pulse that is initialized with the self-similar distribution remains nearly stationary in $\chi_1$-$\chi_2$ space. The part of the distribution that moves is the high-frequency tail, which is a nonlinear effect that the theory does not account for. At $k_{p0}z = 210$, the distribution is distorted due to the transition from vacuum to plasma which occurs at $k_{p0}z=30$, but it quickly returns to the self-similar distribution. The pulse then remains in the self-similar distribution until the end of the simulation at $k_{p0}z=25000$, where it reaches a maximum frequency shift of $k_\delta/k_{\delta0}=20$. Although the pulses in rows (b) and (c) were not initialized to match the self-similar distribution, they appear to approach the self-similar distribution over time. This is because photons travel slower when near the major vertices of their orbits than they do when near the co-vertices, so the rotation of the pulses in rows (b) and (c) is slowing down as they approach the self-similar distribution. To prove this, we can define a polar phase-space coordinate $\varphi$, where $\tan(\varphi) = \chi_2/\chi_1$. The phase space angular velocity is then
\begin{equation}
    \label{angularvelocity}
    \frac{d\varphi}{d\theta} = -\frac{\sqrt{7}}{2}\frac{1+\sin(\psi)\sin(2\varphi)}{\cos(\psi)}\;.
\end{equation}
This angular velocity is smallest in magnitude at $\varphi=-\pi/4$, when the photon is at the major vertex of its trajectory, and it is largest in magnitude at the co-vertex where $\varphi = \pi/4$.

\section{\label{sec:scale} Scaling Laws from the Linear Theory}

With the self-similar solution exhibiting only the $\xi$ contraction and $k$ expansion of accelerating pulses without the phase-space rotation, simple scaling relationships can be obtained between quantities of interest (pulse duration, energy, electric field, etc.) and the frequency shift $k_\delta/k_{\delta0}$ by disregarding the oscillating factors in Eqs.~\eqref{moment2xi},\eqref{moment2xi}, and \eqref{momentxik}.
\begin{table}[htbp]
\caption{\label{tab:scalingx}%
Summary of scaling laws derived from linearized photon kinetic theory. The quantity in the first column scales as $(k_\delta/k_{\delta0})^x$.
}
\begin{ruledtabular}
\begin{tabular}{lc}
\textrm{Quantity} & \textrm{Exponent, $x$} \\
\hline
Electric field amplitude, $E_0$ & $3/4$ \\
Pulse duration, $\sqrt{\langle\Delta\xi^2\rangle}$ & $-1/2$ \\
Bandwidth, $\sqrt{\langle\Delta k^2\rangle}$ & $1/2$ \\
Normalized vector potential, $a_0$ & $-1/4$ \\
Pulse energy, $U$ & $1$ \\
Spatial chirp, $b$ & $1$ \\
\end{tabular}
\end{ruledtabular}
\end{table}

Conveniently, the scaling relationships summarized in Table \ref{tab:scalingx} remain true on average even when the initial pulse does not meet the requirements for self-similar evolution in Eqs.~\eqref{pulsematching}. This is because the quantities of interest for these non-ideal pulses oscillate about the solutions to the self-similar pulse as discussed in Sec. \ref{sec:linPKT}. Furthermore, these oscillations slow down over time, meaning the evolution is dominated by the power law factors for large frequency shifts. The agreement between OSIRIS simulations and the predicted scalings is shown in Fig.~\ref{fig:scaling}.

\begin{figure}[htbp]
    \centering
    \includegraphics[width=0.8\linewidth]{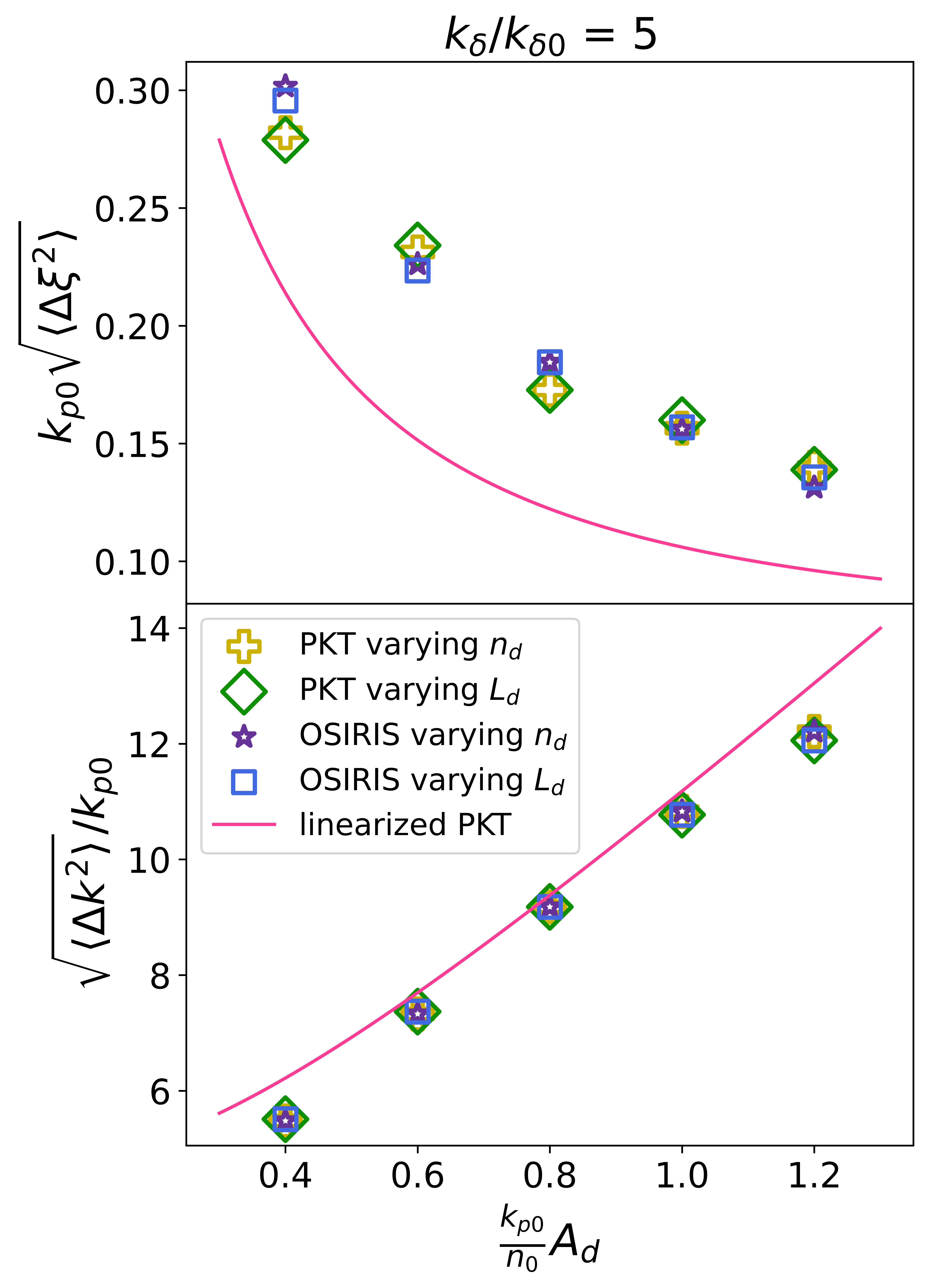}
    \caption{Pulse durations and bandwidths of pulses accelerated with varying areal drive beam density. The durations and bandwidths from the OSIRIS simulations follow the scaling predicted by the linear theory. OSIRIS simulations were ran in 1D, and the PKT simulations used grids of 40 by 40 photons. All simulations used $\tau=1$, $b_0=0$, and $k_{\delta0}=10$. See column D of Table \ref{tab:simparams} for a complete list of simulation parameters.}
    \label{fig:momentsvsA_d}
\end{figure}

The expressions for the pulse duration and bandwidth in Eqs.~\eqref{moment2xi} and \eqref{moment2k} show that, for a constant frequency shift, $\langle\Delta\xi^2\rangle$ decreases as $1/A_d^2$, and $\langle\Delta k^2\rangle$ increases as $A_d^2$ (plus a constant). These scalings are validated using both OSIRIS and PKT simulations. As seen in Fig.~\ref{fig:momentsvsA_d}, the scalings with $A_d$ of pulse durations and bandwidths from both OSIRIS and PKT simulations closely follow the scaling of the linearized solutions. Thus, it can be reasoned that longer, denser drive beams lead to greater pulse compression for the same frequency shift. 

\section{\label{conc}Conclusion}
The linearized photon kinetic theory provides insight into the dynamics at play in the evolution of the phase-space density of the accelerating pulses. The theory predicts that the phase-space density will undergo rotations while contracting in space and expanding in frequency space, causing the pulse to compress indefinitely. The predictions are quantitatively consistent with OSIRIS simulations. There exists a self-similar solution to the linear theory which provides key scaling relationships between pulse parameters and their frequency shift. Particularly, the duration scales with $\omega^{-1/2}$, and the energy scales with $\omega$. PWPA can therefore provide a novel source of ultrafast, high-energy XUV radiation with the potential to scale to arbitrarily low pulse durations.

\section*{Acknowledgments}
This work was funded by NSF grants 2512014 and 2108075.

\appendix
\section{\label{appendixmoments} Moments of the Photon Phase-Space Density}
Statistical moments of the phase-space density in Eq. \eqref{Ndepwpa} can be calculated as follows:
\begin{equation}
    \label{momentdef}
    \langle Q\rangle = \frac{\int QN(\Delta\xi, \Delta k, k_\delta)d\Delta\xi d\Delta k}{\int N(\Delta\xi, \Delta k, k_\delta)d\Delta\xi d\Delta k}\;,
\end{equation}
where $Q$ is a quantity that is a function of the phase-space coordinates $\Delta\xi$ and $\Delta k$. The second moments of the phase-space density are then
\begin{subequations}
    \label{moment2xi}
    \begin{align}
        \langle\Delta\xi^2\rangle &= \frac{C_1 + C_2\sin(\sqrt{7}\theta) - C_3\cos(\sqrt{7}\theta)}{56\tau^2A_d^2k_{\delta0}^2} \frac{k_{\delta0}}{k_\delta}\label{momentxi2} \\
        C_1 &= 128+4\tau^4(2b_0^2 + 3A_dk_{\delta0}b_0 + 2A_d^2k_{\delta0}^2) \label{C1} \\
        C_2 &= \sqrt{7}\tau^4(4A_dk_{\delta0}b_0+3A_d^2k_{\delta0}^2) \label{C2} \\
        C_3 &= 128 + \tau^4(8b_0^2 + 12A_dk_{\delta0}b_0+A_d^2k_{\delta0}^2) \label{C3}      
    \end{align}
\end{subequations}
\begin{subequations}
    \label{moment2k}
    \begin{align}
        \langle\Delta k^2\rangle &= \frac{C_1 - C_4\sin(\sqrt{7}\theta) - C_5\cos(\sqrt{7}\theta)}{56\tau^2}\frac{k_\delta}{k_{\delta0}} \label{momentk2} \\
        C_4 &= \sqrt{7}(48+\tau^4(3b_0^2+4A_dk_{\delta0}b_0)) \label{C4} \\
        C_5 &= 16 + \tau^4(b_0^2 + 12A_dk_{\delta0}b_0 + 8A_d^2k_{\delta0}^2) \label{C5} \\   
    \end{align}
\end{subequations}
\begin{align}
    \langle\Delta\xi\Delta k\rangle =& \frac{-\frac{3}{4}C_1 - \frac{2}{3}(C_2 - C_4)\sin(\sqrt{7}\theta)}{56A_dk_{\delta0}\tau^2} \label{momentxik}\\
    +&\frac{\frac{2}{3}(C_3 + C_5)\cos(\sqrt{7}\theta)}{56A_dk_{\delta0}\tau^2}\;. \notag
\end{align}
Equations \eqref{moment2xi}, \eqref{moment2k}, and \eqref{momentxik} give the spatial variance, spectral variance, and phase-space covariance, respectively, of the accelerating pulses from the linearized theory. The moments are given as functions of the central frequency $k_\delta$, and recall that $\theta=\ln(k_\delta/k_{\delta0})$. Note that the the moments have a factor that oscillates $2\pi$-periodically with $\sqrt{7}\theta$, the spatial variance has a factor that decays with the inverse of the frequency shift, and the spectral variance has a factor that grows linearly with the frequency shift. The oscillations in the covariance have a constant amplitude.

\section{\label{sec:simparams} Simulation Details}
Simulations were performed for a pre-ionized electron plasma in a box with $N_{mesh}$ mesh points and $N_{ppc}$ particles per cell,  with the density profile having a short linear ramp (of length $1$) from $n=0$ to $n=1$ followed by a tailored density profile $n_\delta(z)$. The beam driver had a square profile, and the witness pulse had a Gaussian profile with strength $a_0\equiv eE_0/(m_e\omega_0 c)$, duration $\tau=\tau_{\text{FWHM}}/\sqrt{\ln(2)}$, and initial central wave number $k_{\delta0}$. Table \ref{tab:simparams} summarizes the specific parameters used for each simulation presented in this study.

\begin{table*}[t]
\caption{\label{tab:simparams}%
Parameters used for simulations throughout paper.
The captions of figures that present simulation results state which suite the simulations belong to.
Suite A corresponds to Fig.~\ref{fig:pretty}, Suite B to Figs.~\ref{fig:disptraj}, \ref{fig:wigcontour}, and \ref{fig:moments},
Suite C to Figs.~\ref{fig:wignerchi} and \ref{fig:scaling},
and Suite D to Fig.~\ref{fig:momentsvsA_d}.
}
\begin{center}
\begin{ruledtabular}
\begin{tabular}{ccccc}
\textrm{Simulation Parameter} & \textrm{A} & \textrm{B} & \textrm{C} & \textrm{D} \\
\hline
Dimensionality & quasi-3D & 1D & 1D & 1D \\
Domain Size & $8.5{\times}64$ & $13$ & $15.5$ & $15.25$ \\
$N_\textrm{mesh}$ & $1353{\times}640$ & $10030$ & $8000$ & $20070$ \\
$N_\textrm{ppc}$ & $1{\times}2{\times}16\,(\theta)$ & $1$ & $1$ & $4$ \\
Time Step & $3.70{\times}10^{-3}$ & $8.77{\times}10^{-4}$ & $1.31{\times}10^{-3}$ & $7.60{\times}10^{-4}$ \\
EM Field Solver & Fei & Fei & Fei & Yee \\
$k_{\delta 0}$ & $10$ & $10$ & $3$ & $10$ \\
$\tau$ & $1$ & $1$ & $2.06$ & $1$ \\
$a_0$ & $0.01$ & $0.01$ & $0.01$ & $0.01$ \\
$n_d$ & $0.8$ & $0.8$ & $2.37$ & See Fig.~\ref{fig:momentsvsA_d} \\
$L_d$ & $1$ & $1$ & $0.2$ & See Fig.~\ref{fig:momentsvsA_d} \\
$b_0$ & $0$ & $0$ & See Fig.~\ref{fig:wignerchi} & $0$ \\
\end{tabular}
\end{ruledtabular}
\end{center}
\end{table*}

\end{document}